# Scientific Productivity, Research Funding, Race and Ethnicity

J.S. Yang[1,2], M.W. Vannier[3], F. Wang[4], Y. Deng[4], F.R. Ou[4], J.R. Bennett[1], Y. Liu[4*], G. Wang[1*]

[1]VT-WFU School of Biomedical Engineering and Sciences, Virginia Tech, Blacksburg, USA
[2]School of Mathematical Sciences, Peking University, Beijing, China
[3]Department of Radiology, University of Chicago, Chicago, USA
[4]School of Public Health, China Medical University, Shenyang, China

[*]To whom correspondence should be addressed (YL: cmuliuyang@yahoo.com; GW: ge-wang@ieee.org)

**ABSTRACT**

In a recent study by Ginther *et al.*, the probability of receiving a U.S. National Institutes of Health (NIH) RO1 award was related to the applicant's race/ethnicity. The results indicate black/African-American applicants were 10% less likely than white peers to receive an award, after controlling for background and qualifications. It has generated a widespread debate regarding the unfairness of the NIH grant review process and its correction. In this paper, the work by Ginther *et al.* was augmented by pairing analysis, axiomatically-individualized productivity and normalized funding success measurement. Although there are racial differences in R01 grant success rates, normalized figures of merit for funding success explain the discrepancy. The suggested "*leverage points for policy intervention*" are in question and require deeper and more thorough investigations. Further adjustments in policies to remove racial disparity should be made more systematically for equal opportunity, rather than being limited to the NIH review process.

## 1. BACKGROUND

In a recent study (*D. K. Ginther et al.: "Race, ethnicity, and NIH research awards," Science, 19 August, p. 1015*), the probability of receiving a U.S. National Institutes of Health (NIH) RO1 award was related to the applicant's race/ethnicity. The results indicate black/African-American applicants were 10% less likely than white peers to receive an award, after controlling for background and qualifications, and further suggest "*leverage points for policy intervention*" [1]. These findings have generated a widespread debate regarding the unfairness of the NIH grant review process and its correction. The moral imperative is clear that any hidden racial bias is not to be tolerated, particularly in the NIH funding process. However, the question of whether such a racial bias truly exists requires unbiased, rigorous and systematic evaluation.

NIH director Francis Collins and Deputy Director Lawrence Tabak reiterated that the Ginther study revealed "*from 2000 to 2006, black (1) grant applicants were significantly less likely to receive NIH research funding than were white applicants. The gap in success rates amounted to 10 percentage points, even after controlling for education, country of origin, training, employer characteristics, previous research awards, and publication record (2). Their analysis also showed a gap of 4.2 percentage points for Asians; however, the differences between Asian and white award probabilities were explained by exclusion of noncitizens from the analysis*" [2]. NIH officials admitted "*the gap could also result from 'insidious' bias favoring whites in a peer-review system that supposedly ranks applications only on scientific merit*" [3].

In a Letter to Editor of Science, Dr. Voss expressed uneasiness about proposals which address implications of the Ginther study [4]. He warned that "*disparity-reduction policies represent social experiments with tremendously important consequences, the effects of which could take decades to identify…much of the racial disparity reported could be attributed to black R01 applicants having half the citation count and one-fifth as many last-authored publications as white applicants from similarly ranked institutions. Coupled with the finding that R01s were awarded to highly ranked applications irrespective of* *race, this suggests that R01 disparity is due to lower research success among black applicants rather than to any problems with NIH review*" [4]. In another Letter to Editor, Dr. Erickson pointed out that the citation analysis defined in the Ginther study was not relevant to competitive scientists, the number of citations under consideration should be about 1,000, instead of being about 84, and the number of citations should be normalized to the career length. The opinion was expressed that similarly qualified scientists "*would be equally successful in grant funding, with no disparity for race and ethnicity*" [5].

D. K. Ginther *et al.* wrote a defensive response to these letters. They disagree with Voss about his explanation, because "*there is substantial evidence that affirmative action does not explain the results*" [6]. They found that "*blacks and whites were equally likely to receive tenure at higher education institutions that are research intensive*", and "*a bad match for research careers will have most likely been weeded out earlier*". "*There is a case to be made for positive selection of black scientists – that they are the best of the best – as opposed to being bad matches resulting from affirmative action.*" Also, they disagree with Erickson about the citation issue, because their data included about 300 early-career individuals who had ~1,000 citations, being in the top 1% of the pool. Furthermore, a recent evaluation of the NIH K program [7] showed that awardees published about 10 papers in the 5 years after the award and attracted about 150 citations per person. Furthermore, they did not think that age-normalizing citations would change their results for early-career investigators.

Sherley commented on the view from Tabak and Collins [2], "*the limited public discussion on the possible underlying factors has focused on the NIH review process. Although this is an obvious place to continue the investigation, the explanation may lie elsewhere*" [8]. "*Barriers at the home institution*" were mentioned for "*minority investigators pursuing primarily cancer health disparities research*". For example, "*although NIH requires the writing of minority recruitment plans by its grantee institutions, it currently neither evaluates how nor even whether such plans are implemented.*" Collins and Tabak did not agree with Sherley, "*the plans on all NRSA training grants are rigorously reviewed, and if they are deficient, the grants are not funded until corrective action is taken on the part of the grantee. Awarded training grants that are subsequently submitted for renewal are reviewed for the recruitment plan's results. If the plans are judged ineffective, this assessment affects its likelihood of being funded again*" [9].

Based on the above results and opinions, it is clear that the Ginther study [1] has a tremendous social influence and major policy implications but several countervailing opinions remain unreconciled. Here this issue is re-examined with a new approach and solid data, offering a perspective from paired statistical analysis on NIH funding normalized to individual's scientific productivity. It is the *pairing* and the *normalization* components of the approach that allow deeper insight and more objective conclusions. In the next section, an apparent inconsistency is commented on between the data and conclusions derived by Ginther *et al.*, and an alternative experimental design is proposed. In the third section, an axiomatic approach is defined for quantification of individualized scientific productivity. In the fourth section, the experimental design is described along with key results. In the last section, relevant issues are discussed.

## 2. IS THE GINTHER STUDY SELF-CONSISTENT?

As a general principle, <u>an equitable distribution of research funding should be proportional to each applicant's research capability measured by their scientific productivity</u>. Scholarly publications are a widely-used productivity benchmark, and can be individually quantified by the *h*-index [10] or equivalently number of citations. By this popular metric, the data in the Ginther study [1] does not suggest any significant unfairness in the NIH review process. Specifically, the average white applicant had 78 citations, while the average black applicant had 40 (p. 1018 in [1]). Quite proportionally, the average RO1 success rate was 30%- for white applicants and 15%+ for black (see Fig. 1 in [1]). This citation-based proportionality and the 10% disadvantage noted by Ginther *et al.* seem in contradiction, and motivated the authors to study the issue in more detail.

One potential pitfall of the experimental design by Ginther *et al.* is a sub-optimal use of the Probit model [1], which transforms a continuous sum of weighted variables to a binary outcome (funded or not in this case) via a Gaussian distribution to test the association of race/ethnicity to RO1 success. The variable definitions, inner product (weighted sum), and the Gaussian form of the Probit model could be subject to deficiencies and mismatches, especially for complicated problems. Most remarkably, the criterion for fairness in the funding process has not been well formulated in [1]. Therefore, it seems reasonable to revisit this problem in a more effective fashion. In the following, paired *t*-tests will be used with rigorous matching criteria to spot light any racial difference, and informative features especially funding success normalized by scientific productivity will be extracted to investigate the fairness of the NIH review process.

### 3. HOW TO QUANTIFY INDIVIDUAL SCIENTIFIC PRODUCTIVITY?

While citation count and impact factor are popular measures of publication quality, there is no common agreement on how to quantify relative contributions among co-authors. The number of researchers, publications, and co-authors have all steadily increased over past decades [11]. Consequently, the competition for academic resources has intensified, along with budget squeezes from the current financial crisis. To optimize the resource allocation, individualized assessment of research results is being actively studied [10, 12-17]. However, current indices, such as the numbers of papers and citations, as well as the *h*-factor and its variants [10, 14] have limitations, especially their inability to quantify co-authors' credit shares objectively [18]. Recently, an axiomatic system has been proposed for quantification of co-authors' credits, and the corresponding estimation has been formulated [19], This methodology allows the axiomatically fair measurement of individuals' publication records, avoiding subjective assignment of co-authors' credits using the inflated, fractional or harmonic methods. These findings can be incorporated into existing bibliometric indices for enhancement of their predictive values [20], and has a potential to transform bibliometrics towards a rational framework, providing accurate and practical tools for scientific management.

A recent topic in bibliometrics is the use and extension of the *h*-index [10, 14] for measurement of the productivity and impact of a researcher. While it is increasingly used [21-25], the *h*-index is approximate by definition [26], and subject to various biases [20, 27-35]. A major obstacle to significant improvement of the *h*-index and other popular indices of this type has been the lack of assessment of co-authors' individual contributions.

It is well recognized that the quantification of individual co-authors' credits in a publication is extremely important [12, 13, 15-17]. Current perception of a researcher's qualification relies heavily on either inflated or fractional counting methods [36]; while the former method gives the full credit to any co-author, the latter method distributes an equally divided credit to each co-author. Neither of these methods is ideal because the order or rank of co-authors, and the corresponding authorship, are not used that indicate the relative contributions of co-authors. Generally speaking, the further down the list of co-authors for a publication, the less credit he or she receives; the first and corresponding authors are considered the most prominent.

The harmonic counting method was proposed [36] in order to avoid the equal-share bias of the fractional counting method (a less sophisticated variant was also suggested [17]). While the harmonic counting method does permit equal rankings for subsets of co-authors, let us assume that the order of co-authors' names is consistent with their credit ranking, and that there are $n$ co-authors on a publication whose shares are presented as a vector $\vec{x} = (x_1, x_2, \cdots, x_n)$ ($1 \leq i \leq n$). Then, the *k*-th author contributes $1/k$ as much as the first author. Realistically, there are many possible ratios between the *k*-th and the first author's credits, which may be equal or may be rather small (e.g. cases of data sharing or technical assistance). Hence, the harmonic method has never been used in practice.

There are critical and immediate needs for rigorous quantification of co-authors' credits. The Higher Education Funding Council for England (HEFCE) recently proposed the peer-review system "*Research Excellence Framework (REF)*" [37] that will utilize citation analyses. Nevertheless, HEFCE has admitted that bibliometrics is not "*sufficiently robust*" for assessment of research quality. Thus, it could be prone to misconduct if bibliometric measures are directly used for funding and tenure decisions. For example, a popular Chinese web forum "*New Threads*" [38] discussed several cases of artificially inflated numbers of publications, co-authors, and even *h*-indexes. In the USA, the National Institutes of Health recently adopted enhanced review criteria [39], with mandatory quantification of an investigator's qualification on a 9-point scale (revised from the initially planned 7-point scale); however, the scoring has been largely subjective.

Assume that each publication has $n$ co-authors in $m$ groups ($n \geq m$) where $c_i$ co-authors in the *i*-th group have the same credit $x_i \in \vec{x} = (x_1, x_2, \cdots, x_m)$ ($1 \leq i \leq m$). We postulate the following three axioms:

**Axiom 1:** $x_1 \geq x_2 \geq \cdots \geq x_m > 0$;

**Axiom 2:** $c_1 x_1 + c_2 x_2 + \cdots c_m x_m = 1$;

**Axiom 3:** $\vec{x}$ *is uniformly distributed in the domain defined by Axioms 1 and 2.*

While the first two axioms are self-evident, the third asserts that all the cases permitted by Axioms 1 and 2 are equally possible by the maximum entropy principle [40]. Therefore, the fairest estimation of co-authors' credits must be the expectation of all possible credit vectors. In other words, the *k*-th co-author's credit must be the corresponding elemental mean, which has a closed form expression [19], which is referred to as the *a*-index for its axiomatic foundation.

Naturally, three individualized scientific productivity measures can be defined. First, the productivity measure in terms of journal reputation, or the *Pr*-index, is the sum of the journal impact factors (IF) of one's papers weighted by his/her *a*-indices respectively. Second, the productivity measure in terms of peers' citations, or the *Pc*-index, is the numbers of citations to his/her papers weighted by *a*-indices respectively. While the *Pr*-index is useful for immediate productivity measurement, the *Pc*-index is retrospective and generally more relevant. Finally, the *Pc\*IF* index the sum of the numbers of citations after being individually weighted by both the *a*-index and journal impact factor. When papers are cited, the Pc*IF index credits high-impact journal papers more than low-impact counterparts, as higher-impact papers generally carry tighter relevance or offer stronger support to a citing paper.

# 4. DOES RACIAL BIAS EXIST IN THE NIH REVIEW PROCESS?

## 4.1. Human Subjects

This study targeted the top 92 American medical schools ranked in the 2011 US News and World Report, from which the 31 odd-number-ranked schools were selected for paired analysis (schools were excluded if they did not provide online faculty photos or did not allow 1:2 pairing of black versus white faculty members). Data were gathered from September 1 to 5, 2011 on black and white faculty members in departments of internal medicine, surgery, and basic sciences in the 31 selected schools. White and black/African American faculty members were confirmed by their photos, names, and resumes as needed, and department heads/chairs were excluded. These schools were categorized into three tiers according to their ranking: 1st-31st as the first tier, 33rd-61st as the second tier, and 63rd-91st as the third tier. After 130 black faculty members were found from these schools, 40 black faculty members were randomly selected. With the pairing criteria including the same gender, degree, title, specialty and university, the selected 40 black faculty members were 1:2 paired with white peers, yielding 120 samples as our first pool.

Among the 130 black samples in the initial list, 14 faculty members were funded by NIH during the period from 2008 to 2011. Two of 14 black samples were excluded because of failure in matching with a white faculty. Furthermore, an additional black faculty member was excluded because he only published at conference without any Science Citation Index (SCI) record in this period [41]. Consequently, 11 funded black faculty members were kept. Among them, 10 were from the first tier, and 1 from the second tier. These 11 funded black faculty members were 1:1 paired with white samples who both met the pairing criteria and were funded by NIH in the same period. Consequently, there were 11 pairs of black and white investigators, which is our second pool.

## 4.2. Data Analysis

Using the Web of Knowledge [41], datasets were systematically collected for the two pools of faculty members. Each dataset corresponded to a single black-white combination, and included bibliographic information, such as co-authors, assignment of the corresponding author(s), journal impact factors, and citations 2008-2011. The journal impact factors were obtained from Journal Citation Reports [42].

The *a*-index values were computed using the formula derived by Wang and Yang [19]. In computing *a*-index values, the first author(s) and the corresponding author(s) were treated with equal weights in this context. For the NIH-funded samples, individual numbers of funded proposals and individual funding totals were found via the NIH Reporter system [43].

Our features of interest included the number of journal papers, number of citations, *Pr*-index, *Pc*-index, and *Pc*IF*-index. In addition, for the second pool samples additional features were numbers of NIH funded proposals and NIH funding totals per person and per racial group, respectively.

The paired t-tests were performed using SPSS 13.0 on the datasets from the first and second pools. In the first pool, the average data of two white professors were paired to individual data of the corresponding black professor. The tests were specifically performed by professional rank and school reputation, gender and integrated for racial groups.

## 4.3. Key Results

The scientific productivity was evaluated using the *Pr*-index, *Pc*-index, and *Pc*IF*. Statistical significance levels are indicated by "*" for $p<0.05$ and "**" for $p<0.01$.

Table 1 suggests that higher scientific productivity was positively correlated with more senior professional titles or more prestigious institutional tiers. Furthermore, the analysis shows male investigators were statistically more productive than the female colleagues, and black faculty members statistically less productive than white colleagues. The distribution of professional titles (Full, Associate, and Assistant Professor) for black faculty members was 3:12:25, indicating an imbalance in the higher ranks. Despite that more than a half of the black samples were from first tier institutions, 14 were assistant professors. Thus, the numbers of black associate and full professors were insufficient for us to devise title-specific conclusions with statistical significance.

Table 2 focuses on the scientific productivities of the NIH funded black and white investigators, and indicates similar racial differences in scientific productivity. Although statistical significance cannot be established per professional title due to the limited numbers of samples, the differences between the racial groups are significant in terms of the number of citations and the *Pc*-index. In the following analysis, these scientific productivity measures will serve as the base to evaluate the fairness of the NIH funding process. Note that the racial/ethnic differences in *Pr* and *Pc* (Tables 1 and 2) are consistent with the citation analysis performed in [1].

In Tables 3 and 4, the funding support and the number of funded projects for each racial group were normalized by *Pr*, *Pc* and *Pc*IF* respectively. In addition to the racial difference in the RO1 success rates [1], it can be seen in Tables 3 and 4 that the funding total and the number of funded projects for black NIH investigators were only 46% and 62% of that for whites, respectively. However, when these funding totals and numbers of funded projects were normalized by *Pr*, the ratios between black and white faculty members were narrowed. Furthermore, the normalization by the citation-oriented indices *Pc* and *Pc*IF* indicates that black faculty members had more favorable ratios from 1.06 to 2.00.

|  | Race | Number of Samples | Mean | | | | |
|---|---|---|---|---|---|---|---|
|  |  |  | Mean of Papers | Number of Citations | Pr-index | Pc-index | Pc*IF-index |
| *Full Professor* | Black | 3 | 16.33±17.24 | 120.67±144.36 | 17.62±23.21 | 33.24±50.06 | 130.51±202.80 |
|  | White | 6 | 17.67±22.87 | 197.83±279.04 | 17.49±19.77 | 20.96±26.88 | 260.35±326.53 |
| *Associate Professor* | Black | 12 | 5.83±5.75 | 30.00±37.10 | 4.73±5.25 | 4.69±5.35 | 31.32±42.73 |
|  | White | 24 | 9.08±8.63 | 52.25±55.76 | 5.38±4.55 | 7.78±6.04 | 41.23±58.22 |
| *Assistant Professor* | Black | 25 | 2.44±3.11** | 8.88±20.35* | 1.71±2.17** | 0.86±1.29* | 2.87±5.49* |
|  | White | 50 | 5.18±4.86 | 31.94±52.94 | 6.05±6.42 | 7.05±11.23 | 48.42±107.01 |
| *First Tier (Groups 1-21)* | Black | 21 | 5.19±8.18** | 27.62±63.63* | 5.29±9.92* | 6.09±19.63 | 29.13±82.78 |
|  | White | 42 | 10.02±10.66 | 70.31±118.28 | 9.22±9.38 | 11.07±14.88 | 87.12±168.07 |
| *Second Tier (Groups 22-29)* | Black | 8 | 6.00±6.28 | 36.50±45.26 | 3.41±3.36 | 4.91±6.08 | 24.14±29.35 |
|  | White | 16 | 5.69±5.32 | 26.44±26.85 | 6.20±5.51 | 6.71±5.77 | 37.82±51.48 |
| *Third Tier (Groups 30-40)* | Black | 11 | 2.09±1.81 | 6.55±8.66 | 1.26±1.42 | 0.94±1.38 | 3.12±6.82 |
|  | White | 22 | 3.23±2.79 | 30.09±53.54 | 2.28±2.33 | 4.21±6.10 | 32.22±64.83 |
| *Male* | Black | 22 | 6.14±7.91* | 36.55±65.60 | 4.72±9.17** | 6.60±19.27 | 32.58±81.54* |
|  | White | 44 | 9.68±10.42 | 66.25±111.14 | 8.79±8.82 | 9.93±11.21 | 75.90±135.35 |
| *Female* | Black | 18 | 2.50±4.16 | 7.78±11.79 | 2.69±4.71 | 1.79±2.93 | 6.81±11.68 |
|  | White | 36 | 4.36±4.50 | 31.19±59.12 | 4.16±5.60 | 6.33±12.44 | 45.37±123.49 |
| *Total* | Black | 40 | 4.50±6.68** | 23.60±50.87* | 3.81±7.49** | 4.44±14.48 | 20.98±61.71* |
|  | White | 80 | 7.29±8.63 | 50.48±92.12 | 6.71±7.81 | 8.31±11.77 | 62.16±129.42 |
|  | Ratio | 0.5 | 0.62 | 0.47 | 0.57 | 0.53 | 0.34 |

*Table 1: Scientific productivity measures for black and white faculty members in the first pool.*

| Race | Number of Samples | Mean | | | | |
|---|---|---|---|---|---|---|
|  |  | Number of Papers | Number of Citations | Pr-index | Pc-index | Pc*IF-index |
| *Black* | 11 | 10.45±9.02 | 88.64±98.30* | 11.13±12.47 | 14.96±24.11* | 90.43±124.94 |
| *White* | 11 | 18.64±14.18 | 203.73±189.02 | 18.03±13.24 | 34.39±43.82 | 318.42±474.53 |
| *Ratio* | 1 | 0.56 | 0.44 | 0.62 | 0.44 | 0.28 |

*Table 2: Scientific productivity measures for black and white faculty members in the second pool.*

| Race | Number of Samples | Funding Total | Funding Total Normalized by Pr-index | Funding Total Normalized by Pc-index | Funding Total Normalized by Pc*IF-index |
|---|---|---|---|---|---|
| *Black* | 11 | 20140082 | 164565.69 | 122423.76 | 20247.54 |
| *White* | 11 | 43796537 | 220860.92 | 115781.91 | 12503.74 |
| *Ratio* | 1 | 0.46 | 0.75 | 1.06 | 1.62 |

*Table 3: Ratios between the total funding amount and the accumulated scientific productivity for racial groups (not individuals) in the second pool.*

| Race | Number of Samples | Number of Projects | Number of Projects Normalized by Pr-index | Number of Projects Normalized by Pc-index | Number of Projects Normalized by Pc*IF-index |
|---|---|---|---|---|---|
| **Black** | 11 | 22 | 0.180 | 0.134 | 0.022 |
| **White** | 11 | 37 | 0.187 | 0.098 | 0.011 |
| **Ratio** | 1 | 0.59 | 0.96 | 1.37 | 2.0 |

*Table 4: Ratios between the total number of funded projects and the accumulated scientific productivity for racial groups (not individuals) in the second pool.*

## 5.  DISCUSSION AND CONCLUSION

There are apparent differences in research performance by major racial groups based on individual scientific productivity measures. These findings are consistent with previous reports [1]. The application of the new scientific productivity indices to the racial groups (Tables 1 and 2) clarifies the source of discrepant funding success. When the total grant amounts and the number of funded projects were racial-group-wise normalized by these indices, the NIH review process does not appear biased against black faculty members (Tables 3 and 4). Specifically, the funding total and the number of funded projects for black NIH investigators were respectively only 46% and 62% of that for white peers. However, when these funding totals and the number of funded projects were normalized by *Pr*, the ratios between black and white faculty members neared parity. Furthermore, the normalization by the citation-oriented indices *Pc* and Pc*IF indicates that black researchers are not in a disadvantageous position.

The axiomatically derived *a*-, *Pr*-, *Pc*-, and *Pc*IF*- indices individualize credits (journal impact factor, number of citations, or both) for coauthored papers or other forms of joint teamwork. These metrics apportion an integrated contribution most equitably among researchers so that credit can be quantitatively shared for team science activity. All figures of merit including axiomatically derived ones have limitations but assessment of scientific productivity and research potential should be done to be commensurate with individual contributions. Originality, novelty, healthcare impact, and peers' perception are all critical facets of the assessment. The axiomatic approach is advantageous due to rigor and objectivity, should be positively correlated to the other quantitative and qualitative criteria, and could be helpful in the NIH funding process to quantify achievements, detect disparity, and facilitate management. In particular, such tools could aid streamlining and monitoring of peer-review and research execution.

The key results achieved statistical significance, when subjected to paired analysis capable of sensing differences with adequate specificity and sensitivity. There is potential for the axiomatic approach to produce more comprehensive results with expansion of the sample size. The databases construction used in this study took our 10 students' efforts over about three months, and yet cannot be compared with that used in the Ginther study in terms of sample size (The Ginther study was based on a much larger sample size, "*this sample included 83,188 observations with non-missing data for the explanatory variables*" [1]). On the other hand, if there were detailed information on educational background, training, prior awards, and related variables, pairing of black and white investigators could become impossible in many cases. In this study, the critical abstraction across various groups has been axiomatically-formulated scientific productivity and accordingly-defined funding normalization. This perspective allows us to evaluate the fairness of the NIH review process in a more straightforward way.

The limitations of the current study are multiple, and have compromised the results to different degrees. The research disciplines, specific institutes, other grant mechanisms (e.g., P and K awards) were not separately considered. The prior training (T and K awards), longitudinal trends, and review process changes were not analyzed. When the samples were selected, the unavailability of some faculty photos was a difficulty. Since the number of white faculty members is large, it was hoped to use more white samples for a better representation. However, the pairing criteria prevented us from including white faculty members beyond the 1:2 and 1:1 ratios for the first and second pools, respectively. The existing online searching systems do not support the computation of the axiomatic indices. The tedious data entry and analysis tasks are error-prone. Cross validation steps were performed to produce data up to a high standard. Ideally, an automated exclusive study using the axiomatic approach should be performed to generate the highest possible statistical confidence. In this regard, the Ginther study is a model.

Axiomatically-oriented bibliometrics employs value theory to address the basic question of how, why and what value is ascribed to an individual's scientific work. Cultural, sociological, geographical, psychological, economical, physical, computational, and other factors influence the results. In the 19th century, Adam Smith asserted that the amount of labor put into a physical product determined its exchange value. The concept was refined by others, including Karl Marx, John Keynes, and the Chicago school of economics, but the valuation of an intellectual product is much more challenging. Although extensive studies have been done on this topic, including citation analysis, there has been no reliable means to value individual credits in teamwork or joint publications. An axiomatic theory for individualized quantification of scientific productivity [19] introduced to address this need, was used in this racial disparity study. In the future, major search engines such as Web of Science and Google Scholar may implement the *Pr*-, *Pc*- and *Pc\*IF*-indices to augment individual productivity assessments.

Although the NIH review process endeavors to be racially fair, it is not perfect in all aspects. How to evaluate and optimize the NIH funding process has been a hot topic [44]. The NIH Grant Productivity Metrics and Peer Review Scores Online Resource [45, 46] stimulates hypotheses that can be tested using the axiomatic indices. For example, will new investigators be more influential than senior researchers? Will large grant mechanisms such as U01 and P41 be more productive than R01 and R21? Will renewed projects be more cost-effective than initially funded projects? Although any bibliometric measures are subject to inter-specialty fluctuations, some of commonly interested problems can be studied using the axiomatic indices with the same individual or team as its own control.

In conclusion, the NIH grant racial disparity study of Ginther et al. [1] was augmented by a pairing-based axiomatically-individualized productivity and normalized funding success measurement trial. Although there are racial differences in R01 grant success rates [1], normalized figures of merit for funding success explain the discrepancy. The suggested "*leverage points for policy intervention*" [1] are in question and require deeper and more thorough investigations, given the important social complications of this sensitive issue. Further adjustments in policies to remove racial disparity should be made more systematically for equal opportunity, rather than being limited to the NIH review process.

**Acknowledgment**

The authors are grateful for diligent data entry and analysis work performed by our research assistants especially W.Y. Wang, S.H. Tong, Y.Y. Gao, and G.C. Liu.

# References

1. Ginther, D.K., et al., Race, Ethnicity, and NIH Research Awards. Science, 2011. 333(6045): p. 1015-1019.
2. Tabak, L.A. and F.S. Collins, Weaving a Richer Tapestry in Biomedical Science. Science, 2011. 333(6045): p. 940-941.
3. Kaiser, J., NIH Uncovers Racial Disparity in Grant Awards. Science, 2011. 333(6045): p. 925-926.
4. Voss, J.L., Race Disparity in Grants: Empirical Solutions Vital. Science, 2011. 334(6058): p. 899-899.
5. Erickson, H.P., Race Disparity in Grants: Check the Citations. Science, 2011. 334(6058): p. 899-899.
6. Ginther, D.K., et al., Race Disparity in Grants: Empirical Solutions Vital Response. Science, 2011. 334(6058): p. 899-U154.
7. http://grants.nih.gov/training/K_Awards_Evaluation_FinalReport_20110901.pdf.
8. Sherley, J.L., Race Disparity in Grants: Oversight at Home. Science, 2011. 334(6058).
9. Collins, F.S. and L.A. Tabak, Race Disparity in Grants: Oversight at Home Response. Science, 2011. 334(6058).
10. Hirsch, J.E., An index to quantify an individual's scientific research output. Proc Natl Acad Sci U S A, 2005. 102(46): p. 16569-72.
11. Greene, M., The demise of the lone author. Nature, 2007. 450(7173): p. 1165.
12. Foulkes, W. and N. Neylon, Redefining authorship. Relative contribution should be given after each author's name. BMJ, 1996. 312(7043): p. 1423.
13. Campbell, P., Policy on papers' contributors. Nature, 1999. 399(6735): p. 393.
14. Hirsch, J.E., Does the H index have predictive power? Proc Natl Acad Sci U S A, 2007. 104(49): p. 19193-8.
15. Anonymous, Who is accountable? Nature, 2007. 450(7166): p. 1.
16. Ball, P., A longer paper gathers more citations. Nature, 2008. 455(7211): p. 274-5.
17. Zhang, C.T., A proposal for calculating weighted citations based on author rank. EMBO Rep, 2009. 10(5): p. 416-7.
18. Williamson, J.R., My h-index turns 40: my midlife crisis of impact. ACS Chem Biol, 2009. 4(5): p. 311-3.
19. Wang, G. and J.S. Yang, Axiomatic quantification of co-authors' relative contributions. arXiv.org, 2010: p. arXiv:1003.3362v1 [stat.AP].
20. Bornmann, L. and H.D. Daniel, The state of h index research. Is the h index the ideal way to measure research performance? EMBO Rep, 2009. 10(1): p. 2-6.
21. Ball, P., Index aims for fair ranking of scientists. Nature, 2005. 436(7053): p. 900.
22. Ball, P., Achievement index climbs the ranks. Nature, 2007. 448(7155): p. 737.
23. Kinney, A.L., National scientific facilities and their science impact on nonbiomedical research. Proc Natl Acad Sci U S A, 2007. 104(46): p. 17943-7.
24. Pilc, A., The use of citation indicators to identify and support high-quality research in Poland. Arch Immunol Ther Exp (Warsz), 2008. 56(6): p. 381-4.
25. Sebire, N.J., H-index and impact factors: assessing the clinical impact of researchers and specialist journals. Ultrasound Obstet Gynecol, 2008. 32(7): p. 843-5.
26. Dodson, M.V., Citation analysis: Maintenance of h-index and use of e-index. Biochem Biophys Res Commun, 2009. 387(4): p. 625-6.
27. Lehmann, S., A.D. Jackson, and B.E. Lautrup, Measures for measures. Nature, 2006. 444(7122): p. 1003-4.
28. Jeang, K.T., Impact factor, H index, peer comparisons, and Retrovirology: is it time to individualize citation metrics? Retrovirology, 2007. 4: p. 42.
29. Kelly, C.D. and M.D. Jennions, H-index: age and sex make it unreliable. Nature, 2007. 449(7161): p. 403.
30. Wendl, M.C., H-index: however ranked, citations need context. Nature, 2007. 449(7161): p. 403.

31. *Engqvist, L. and J.G. Frommen, The h-index and self-citations. Trends Ecol Evol, 2008. 23(5): p. 250-2.*
32. *Mishra, D.C., Citations: rankings weigh against developing nations. Nature, 2008. 451(7176): p. 244.*
33. *Radicchi, F., S. Fortunato, and C. Castellano, Universality of citation distributions: toward an objective measure of scientific impact. Proc Natl Acad Sci U S A, 2008. 105(45): p. 17268-72.*
34. *Todd, P.A. and R.J. Ladle, Citations: poor practices by authors reduce their value. Nature, 2008. 451(7176): p. 244.*
35. *Baldock, C., R. Ma, and C.G. Orton, Point/counterpoint. The h index is the best measure of a scientist's research productivity. Med Phys, 2009. 36(4): p. 1043-5.*
36. *Hagen, N.T., Harmonic allocation of authorship credit: source-level correction of bibliometric bias assures accurate publication and citation analysis. PLoS One, 2008. 3(12): p. e4021.*
37. *http://www.nature.com/news/2009/090923/full/news.2009.933.html.*
38. *http://www.xys.org/new.html.*
39. *http://grants.nih.gov/grants/guide/notice-files/NOT-OD-09-024.html.*
40. *Jaynes, E.T., On the Rationale of Maximum-Entropy Methods. Proceedings of the Ieee, 1982. 70(9): p. 939-952.*
41. *http://sub3.webofknowledge.com.*
42. *http://thomsonreuters.com/products_services/science/science_products/a-z/journal_citation_reports.*
43. *http://projectreporter.nih.gov/reporter.cfm.*
44. *https://loop.nigms.nih.gov/index.php/category/peer-review.*
45. *https://loop.nigms.nih.gov/index.php/2011/06/02/productivity-metrics-and-peer-review-scores.*
46. *https://loop.nigms.nih.gov/index.php/2011/06/10/productivity-metrics-and-peer-review-scores-continued.*